\documentclass[12pt]{article}
\usepackage{amssymb}
\usepackage[dvips]{epsfig}

\begin{document}

\author{A. de Souza Dutra$^{a,b}$\thanks{%
E-mail: dutra@feg.unesp.br}, A.C. Amaro de Faria Jr.$^{b}$ and M. Hott$^{b}$%
\thanks{%
e-mail: hott@feg.unesp.br} \\
%EndAName
\\
$^{a}$Abdus Salam ICTP, Strada Costiera 11, Trieste, I-34100 Italy.\\
$^{b}$UNESP-Campus de Guaratinguet\'{a}-DFQ\thanks{
Permanent Institution}\\
Departmento de F\'{\i}sica e Qu\'{\i}mica\\
12516-410 Guaratinguet\'{a} SP Brasil}
\title{{\LARGE Degenerate and critical Bloch branes}}
\maketitle

\begin{abstract}
In the last few years a number of works reported the appearance of thick
branes with internal structure, induced by the parameter which controls the
interaction between two scalar fields coupled to gravity in (4,1) dimensions
in warped space-time with one extra dimension. Here we show that one can
implement the control over the brane thickness without needing to change the
potential parameter. On the contrary, this is going to be done by means of
the variation of a parameter associated to the domain wall degeneracy. We
also report the existence of novel and qualitatively different solutions for
a critical value of the degeneracy parameter, which could be called critical
Bloch branes. \newline
\end{abstract}

\newpage

\section{\protect\smallskip Introduction}

The problem of the so called thick branes have received a considerable
amount of attention along the last years \cite{campos}-\cite{giovannini2}.
As observed by Campos \cite{campos} a few years ago, some kinds of two
interacting scalar fields potentials can be used in order to describe the
splitting of thick branes due to a first-order phase transition in a warped
geometry. In that work, Campos discussed the effect by studying a model
without supersymmetry. In a recent work, Bazeia and Gomes \cite{gomes}
discussed the appearance of thick branes by using a two interacting scalar
fields model that can be naturally incorporated in supergravity, what was
also done by Eto and Sakai \cite{eto}. In our work, by using the very same
model discussed by Bazeia and Gomes, we show that the reported Bloch branes
\cite{gomes} have, in fact, more general soliton solutions and, as a
consequence, the resulting brane can have a much richer structure as, for
instance, a degeneracy controlling parameter \cite{eto,shiffman}. It is
shown that for a convenient choice of this parameter a double wall structure
and a corresponding thicker Bloch Brane, what we call \textit{degenerate
Bloch branes} shows up. It is important to remark that both in the case of
Campos as in the Bazeia and Gomes, the splitting of the thick branes is
controlled by the potential parameters. Here, instead, the splitting is
controlled by means of a parameter which is not present in the Lagrangian
density, on the contrary it appears in the solutions as a shape controlling
parameter. In fact, as asserted in above, there is a degeneracy in the
solution because, in spite of the value of this parameter the energy of the
field configuration is precisely the same. In view of this we call it a
degeneracy parameter \cite{shiffman}. In fact, Bazeia and collaborators \cite%
{bazeiaPRL03} have introduced a one scalar field model with similar
properties like the existence of double-walls solutions and thick branes
\cite{furtado} but, once again, the brane thickness control is done through
a fine tuning of the potential self-interaction parameters themselves. Here
the idea is to get a more robust way of controlling the thickness, and the
correspondent distance between the walls whereas preserving a supersymmetric
structure. Finally we introduce a special solution, at the critical value of
the degeneracy parameter, which presents a quite different and interesting
behavior for the brane.

On the other hand finding exact classical solutions, particularly solitons,
is one of the problems on nonlinear models with interacting fields \cite%
{Rajaraman}-\cite{lorentzbr}. As pointed out by R. Rajaraman and E. Weinberg
\cite{rajaramanweinberg}, in such nonlinear models more than one
time-independent classical solution can exist and each one of them
corresponds to a different family of quantum states, which come into play
when one performs a perturbation around those classical solutions.

In order to deal with the systems we are going to work here, it is common to
use the so called \textit{trial orbits method} \cite{rajaramanmethod}, which
is a very powerful one presented for finding exact soliton solutions for
non-linear second-order differential equations of models with two
interacting relativistic scalar fields in 1+1 dimensions, and it is model
independent. A couple of years ago one of us presented a method for finding
additional soliton solutions for those special cases whose soliton solutions
are the BPS ones \cite{PLB05} and in the last year that approach was
extended, allowing more general models \cite{PLB06,boya,lorentzbr}. This
last approach is the one we will use along this manuscript. As a
consequence, we present more general soliton solutions and how they are
intrinsically related to degenerate and critical Bloch branes. Furthermore,
once we have found more general solutions which engenders thicker branes, we
discuss the influence of those solutions in the warp factor and in the
fluctuation of the metric around those classical solutions. We also compare
our results with those obtained in \cite{gomes}.

This work is organized as follows: In the second section we present the
model we are going to work with and review the approach introduced in the
reference \cite{PLB05} to find classical soliton solutions. In that section
we also obtain the warp factor in a general form. In the third section, a
variety of soliton solutions that have been found up to now, are constructed
by using the method of the second section; we also present the warp factor
for each set of solutions. The following section is devoted to discuss the
stability and the zero modes for each set of soliton solutions in the
context of the brane worlds scenario. Finally, we address final comments on
the soliton solutions and their consequences and applicability in the brane
world scenario.

\section{Gravity coupled to two interacting scalar fields: analytical
solutions}

The action we are going to work with, is the one of a five-dimensional
gravity coupled to two interacting real scalar fields, which can be
represented by \cite{dewolfe, dewolfe1}
\begin{equation}
S=\int d^{4}x\,dr\sqrt{\left\vert g\right\vert }\left[ -\frac{1}{4}R+\frac{1%
}{2}\left( \partial _{\mu }\phi \,\partial ^{\mu }\phi +\partial _{\mu }\chi
\,\partial ^{\mu }\chi \right) -V\left( \phi ,\chi \right) \right] ,
\end{equation}

\noindent where $g\equiv \det \left( g_{ab}\right) $ and for granting that
the four-dimensional space has the Poincar\'{e} invariance, it is usually
used that
\begin{equation}
ds^{2}=g_{ab}\,dx^{a}dx^{b}=e^{2\,A\left( r\right) }\eta _{\mu \nu }dx^{\mu
}dx^{\nu }-dr^{2};\,a,b=0,...,4,
\end{equation}

\noindent where $r$ is the extra dimension, $\eta _{\mu \nu }$ the usual
Minkowski metric and $e^{2\,A\left( r\right) }$ is the so called warp
factor. A usual hypothesis is that the warp factor depends only on the extra
dimension $r$. Besides, one can also assume that the scalar fields depend
only of the extra dimension $r$. Under these assumptions, one can determine
the resulting equations of motion for the above system as \cite{gomes,
dewolfe, dewolfe1}
\begin{equation}
\frac{d^{2}\phi }{dr^{2}}+4\,\frac{dA}{dr}\,\frac{d\phi }{dr}=\frac{\partial
V\left( \phi ,\chi \right) }{\partial \phi },\,\,\frac{d^{2}\chi }{dr^{2}}%
+4\,\frac{dA}{dr}\,\frac{d\chi }{dr}=\frac{\partial V\left( \phi ,\chi
\right) }{\partial \chi },
\end{equation}

\noindent and
\begin{equation}
\frac{d^{2}A}{dr^{2}}=-\frac{2}{3}\left[ \left( \frac{d\phi }{dr}\right)
^{2}+\left( \frac{d\chi }{dr}\right) ^{2}\right] ,\,\,\left( \frac{dA}{dr}%
\right) ^{2}=\frac{1}{6}\left[ \left( \frac{d\phi }{dr}\right) ^{2}+\left(
\frac{d\chi }{dr}\right) ^{2}\right] -\frac{1}{3}V\left( \phi ,\chi \right) .
\end{equation}

As can be demonstrated \cite{dewolfe,gomes}, the above set of second-order
non-linear coupled equations has another set of first-order differential
equations which shares solutions with it, and that is given by
\begin{equation}
\frac{d\phi }{dr}=\frac{\partial W\left( \phi ,\chi \right) }{\partial \phi }%
,\,\,\frac{d\chi }{dr}=\frac{\partial W\left( \phi ,\chi \right) }{\partial
\chi },\,\,\frac{dA}{dr}=-\frac{2}{3}W\left( \phi ,\chi \right) ,
\label{1storder}
\end{equation}

\noindent provided that the potential $V\left( \phi ,\chi \right) $ is
restricted to be of a given class of potentials which can be written in
terms of a kind of superpotential as
\begin{equation}
V\left( \phi ,\chi \right) =\frac{1}{2}\left[ \left( \frac{\partial W\left(
\phi ,\chi \right) }{\partial \phi }\right) ^{2}+\left( \frac{\partial
W\left( \phi ,\chi \right) }{\partial \chi }\right) ^{2}\right] -\frac{4}{3}%
W\left( \phi ,\chi \right) ^{2}.  \label{pot}
\end{equation}

\noindent Note that, once one has solutions of the first two first-order
equations for the interacting scalar fields, it becomes a simple task of
integration to get $A\left( r\right) $ and, as a consequence, determine the
warp factor. Now, in order to go further on the analysis it is important to
work with a concrete example which, when available, should be one with exact
analytical solutions \cite{gremm}. So, we will work with a superpotential
equivalent to that used by Bazeia and collaborators \cite{gomes}. The idea
is to show that there are other solutions which were not analyzed in
reference \cite{gomes}, and which present quite interesting features.
Particularly, we will show that some of them allow one to control the
behavior of the warp factor without performing the kind of restriction over
the potential parameters as the one did in \cite{gomes}. The superpotential
we are going to work with along this manuscript is
\begin{equation}
W(\phi ,\chi )=\phi \left[ \lambda \left( \frac{\phi ^{2}}{3}-a^{2}\right)
+\mu \,\chi ^{2}\right] ,  \label{superpot}
\end{equation}

\noindent which becomes equal to that considered in \cite{gomes} by choosing
$a=1$, $\lambda =-1$ and $\mu =-\,r$. From now on, in order to solve the
first-order differential equations, we follow the method of reference \cite%
{PLB05} instead of applying the usual trial orbits method \cite%
{rajaramanmethod}, \cite{bazeiareview}. For this we note that it is possible
to write the relation $d\phi /W_{\phi }=dr=d\chi /W_{\chi }$, where the
differential element $dr$ is a kind of invariant. Thus, one is lead to

\begin{equation}
\frac{d\phi }{d\chi }=\frac{W_{\phi }}{W_{\chi }}.  \label{eqm}
\end{equation}
This is in general a nonlinear differential equation relating the scalar
fields of the model. If one is able to solve it completely for a given
model, the function $\phi \left( \chi \right) $ (in fact, it will be the
equation for a generic orbit) can be used to eliminate one of the fields,
rendering the first-order differential equations uncoupled and equivalent to
a single one. Finally, the resulting uncoupled first-order nonlinear
equation can be solved in general, even if numerically. By substituting the
derivatives of the superpotential (\ref{superpot}) with respect to the
fields in (\ref{eqm}) we have
\begin{equation}
\frac{d\phi }{d\chi }=\frac{\lambda (\phi ^{2}-a^{2})+\mu ~\chi ^{2}}{2~\mu
~\phi ~\chi },  \label{eqm2}
\end{equation}
which can be rewritten as a linear differential equation,
\begin{equation}
\frac{d\rho }{d\chi }-\frac{\lambda }{\mu ~\chi }=\chi ,  \label{eqm3}
\end{equation}
by the redefinition of the fields, $\rho =\phi ^{2}-a^{2}$. Now, the general
solutions are easily obtained as
\begin{equation}
\rho (\chi )=\phi ^{2}-a^{2}=c_{0}~\chi ^{\lambda /\mu }-\frac{\mu }{\lambda
-2\mu }~\chi ^{2},\hspace{0.5in}\mathrm{for\ \ }\lambda \neq 2\mu ,
\label{sol1eqm}
\end{equation}
\noindent and
\begin{equation}
\rho (\chi )=\phi ^{2}-a^{2}=\chi ^{2}[\ln (\chi )+c_{1}],\hspace{0.5in}%
\mathrm{for~\ }\lambda =2\mu ,  \label{sol2eqm}
\end{equation}
where $c_{0}$ and $c_{1}$ are arbitrary integration constants. We substitute
the above solutions, for instance, in the differential equation for the $%
\chi $ field, obtaining the following first-order differential equations for
the field $\chi (r)$%
\begin{equation}
\frac{d\chi }{dr}=\pm \,2\,\mu \chi \sqrt{\,a^{2}+c_{0}\,\chi ^{\lambda /\mu
}-\frac{\mu }{\lambda -2\mu }~\chi ^{2}}\,,\quad \,\,\lambda \neq 2\mu ,
\label{eqdchi1}
\end{equation}

\noindent and
\begin{equation}
\frac{d\chi }{dr}=\pm \,2\,\mu \chi \sqrt{\,a^{2}+\chi ^{2}[\ln (\chi
)+c_{1}]}\,,\quad \,\,\lambda =2\mu .  \label{eqdchi2}
\end{equation}
As a matter of fact, in general, an explicit solution for each one of the
above equations can not be obtained, but one can verify numerically that the
solutions belong to the same classes, and some of those classes of solutions
can be written in terms of analytical elementary functions. In those last
cases one is able to obtain the several types of soliton solutions we
discuss in the next section.

Now, let us discuss a bit on the form of the warp factor $e^{2\,A\left(
r\right) }$, for which it is necessary to compute the function $A\left(
r\right) $ by integrating the second equation in (5). Once we will consider
many situations, some of them with expressions much more involved than those
studied in \cite{gomes}, it should be very convenient to express it in a
general form in terms of the field itself, instead of a function of the
spatial variable. Furthermore, that would allows one to make qualitative
considerations about the behavior of the warp factor. In order to put this
idea in a concrete form, we will use the orbit equations (\ref{sol1eqm}, \ref%
{sol2eqm}) and manipulate the equation for $A\left( r\right) $ in order to
write it in terms of the field $\chi \left( r\right) $. For this we start by
noting that after eliminating the dependence of $A\left( \phi ,\chi \right) $
by using the orbit equation ($\phi \equiv \phi \left( \chi \right) $), one
obtains
\begin{equation}
\frac{dA\left( r\right) }{dr}=\frac{dA\left( \chi \right) }{d\chi }\,\frac{%
d\chi }{dr}=\frac{dA\left( \chi \right) }{d\chi }\,\frac{\partial W\left(
\phi \left( \chi \right) ,\chi \right) }{\partial \chi }=-\frac{2}{3}W\left(
\phi \left( \chi \right) ,\chi \right) ,
\end{equation}

\noindent which leads to
\begin{equation}
\frac{dA\left( \chi \right) }{d\chi }=-\frac{2}{3}\frac{W\left( \phi \left(
\chi \right) ,\chi \right) }{W_{\chi }\left( \phi \left( \chi \right) ,\chi
\right) },
\end{equation}

\noindent where $W_{\chi }\left( \phi \left( \chi \right) ,\chi \right)
\equiv \frac{\partial W\left( \phi \left( \chi \right) ,\chi \right) }{%
\partial \chi }$. Now, substituting the superpotential of the model under
analysis we get
\begin{equation}
\frac{dA\left( \chi \right) }{d\chi }=-\,\frac{1}{3\,\mu \,\chi }\left[
\lambda \left( \frac{\phi \left( \chi \right) ^{2}}{3}-a^{2}\right) +\mu
\,\chi ^{2}\right] ,
\end{equation}

\noindent which after some simple manipulations using the orbit equation
conduces to
\begin{equation}
\frac{dA\left( \chi \right) }{d\chi }=\left( \frac{2\,\lambda \,a^{2}}{%
9\,\mu }\right) \chi ^{-1}-\frac{2}{9}\left( \frac{\lambda -3\,\mu }{\lambda
-2\,\mu }\right) \chi -\left( \frac{\lambda \,c_{0}}{9\,\mu }\right) \chi
^{\left( \frac{\lambda }{\mu }-1\right) },\hspace{0.5in}\mathrm{for\ \ }%
\lambda \neq 2\mu ,
\end{equation}

\noindent and
\begin{equation}
\frac{dA\left( \chi \right) }{d\chi }=\left( \frac{2\,\lambda \,a^{2}}{%
9\,\mu }\right) \chi ^{-1}-\frac{\left( 3\,\mu +\lambda \,c_{1}\right) }{%
9\,\mu }\,\chi -\frac{1}{3\,\mu }\,\chi \,\ln \left( \chi \right) ,\hspace{%
0.5in}\mathrm{for\ \ }\lambda =2\mu .
\end{equation}

\noindent Finally we can perform the integration over the field $\chi $,
obtaining
\begin{equation}
A\left( \chi \right) =\alpha _{0}+\left( \frac{2\,a^{2}}{9\,}\right) \ln
\left( \chi \right) -\frac{1}{9}\left( \frac{\lambda -3\,\mu }{\lambda
-2\,\mu }\right) \chi ^{2}-\left( \frac{\,c_{0}}{9}\right) \chi ^{\left(
\frac{\lambda }{\mu }\right) },\hspace{0.5in}\mathrm{for\ \ }\lambda \neq
2\mu ,
\end{equation}

\noindent and
\begin{equation}
A\left( \chi \right) =\alpha _{1}+\left( \frac{2\,a^{2}}{9\,}\right) \ln
\left( \chi \right) -\frac{\left( 3\,\mu +\lambda \,c_{1}\right) }{18\,\mu }%
\,\chi ^{2}-\frac{1}{6\,\mu }\,\chi ^{2}\,\left( \ln \left( \chi \right) -%
\frac{1}{2}\right) ,\hspace{0.5in}\mathrm{for\ \ }\lambda = 2\mu ,
\end{equation}

\noindent where $\alpha _{0}$ and $\alpha _{1}$ are arbitrary integration
constants, which are going to be chosen to ensure that $A\left( r=0\right)
=0 $. It is important to remark that the above solutions are completely
general for this model and, as a consequence, can be used to get the warp
factor for an arbitrary choice of the potential parameters, even for the
cases where its solution can not be obtained through analytical elementary
functions. The above general approach can be checked, for instance with case
studied by Bazeia and Gomes \cite{gomes}, for which
\begin{equation}
\chi \left( y\right) =\pm \sqrt{\frac{1}{s}-2}\,\,\mathrm{sech}\left(
2\,s\,y\right) \mathrm{,}
\end{equation}

\noindent and, from above, one obtains
\begin{equation}
A\left( y\right) =\frac{1}{9\,s}\left[ \left( 1-3\,s\right) \right] \tanh
\left( 2\,s\,y\right) ^{2}-2\,\ln \left( \cosh \left( 2\,s\,y\right) \right)
,  \label{warpbazeia}
\end{equation}

\noindent where we have used the original variables and parameters defined
in \cite{gomes}. This is precisely the result obtained in that work through
direct integration in the spatial variable.

From the expressions obtained above, one can clearly see that the behavior
of the warp factor is very sensitive to the one of the field $\chi $%
.\thinspace For instance, when $\chi \left( r\right) $ changes very slowly
in a given region, so it will happens with $A\left( r\right) $. In certain
manner, one can guess the behavior of the warp factor, simply by observing
that for $\chi $.

\section{Soliton solutions and their warp factors}

In this section we explore in some extent the solutions for the equation (%
\ref{eqdchi1}). This is done by presenting those resulting soliton solutions
for the model under consideration and also by obtaining explicitly the warp
factor for each set of soliton solutions in the brane scenario under
analysis. Finally we compare our results with those offered in \cite{gomes}.

Before proceeding with this program, we would like to stress that the model
we are working with admits a particular set of solutions which can not be
obtained from the method described in the previous section. That set of
classical solutions, which could be called isolated solutions because is
characterized by $\bar{\chi}_{I}(r)=0$, such that there is no sense in
writing the differential equation (\ref{eqm2}) for this case and,
consequently, do not furnishes any internal structure for the brane \cite%
{gomes}. Even though, the system admits a soliton solution given by $\bar{%
\phi}_{I}(r)=\pm a~\tanh (\lambda \,a\,r)$, where the (lower) upper sign
refers to a (anti-)kink solution. We will not consider this case in detail,
once it is effectively a one field model, and we are primarily interested in
the two fields nontrivial solutions.

\subsection{Bloch walls}

The usual set of solutions, baptized as Bloch wall in \cite{gomes}, can be
obtained by means of the method described in the previous section. It is
obtained when we take $c_{0}=0$ in the expression (\ref{eqdchi1}). In this
case that equation can be solved analytically for any value of $\lambda $
and $\mu $, provided that $\lambda >2\mu $ in order to keep the solution
real. In this case we get the following solution for $\chi (r)$%
\begin{equation}
\chi _{BW}(r)=a\sqrt{\frac{\lambda -2\mu }{\mu }}\mathrm{sech}(2\,\mu
\,a\,r).  \label{chi2a}
\end{equation}%
One can observe that this solution vanishes when $x\rightarrow \pm \infty $.
The corresponding kink-like solution for the field $\phi$, is given by
\begin{equation}
\phi _{BW}(r)=\pm a~\tanh (2\;\mu \;a\;r),  \label{phi2a}
\end{equation}%
which connect the vacua of the potential. In this case the warp factor of
the configuration is the one presented in the previous section in equation (%
\ref{warpbazeia}). We call this type of domain wall as $BW$ domain wall to
distinguish it from other types of domain wall solutions we are going to
present for this model.

At this point it is interesting to note that, some of the solutions we are
going to explore in the next sections are, in fact, in a different range of
the potential parameters as compared with the ones considered in \cite{gomes}%
. For instance, when we consider the $\lambda = \mu$, in terms of the
parameters used by Bazeia and Gomes, this case would corresponds to consider
$r=1$. However the range of validity of the solutions used by them is $0 < r
\leq \frac{1}{2}$.

\subsection{Degenerate Bloch walls}

Others soliton solutions can be found when one considers the integration
constant $c_{0}\neq 0$. It was found in the reference \cite{PLB05} that at
least in three particular cases the equation (\ref{eqdchi1}) can be solved
analytically. For $c_{0}<-2$ and $\lambda =\mu $ it was found that the
solutions for the $\chi (r)$ field are lump-like solutions, which vanishes
when $r\rightarrow \pm \infty $. On its turn, the field $\phi (r)$ exhibits
a kink-like profile.

These classical solutions can be written as
\begin{equation}
\tilde{\chi}_{DBW}^{\left( 1\right) }(r)=\frac{2a}{\sqrt{c_{0}^{2}-4}\cosh
(2\,\mu \,a\,r)-c_{0}},\hspace{0.5in}\mathrm{for}\hspace{0.5in}\lambda =\mu
,~c_{0}<-2,  \label{chi2atil1}
\end{equation}

\noindent and
\begin{equation}
\tilde{\phi}_{DBW}^{\left( 1\right) }(r)=a\frac{\sqrt{c_{0}^{2}-4}\sinh
(2\,\mu \,a\,r)}{\sqrt{c_{0}^{2}-4}\cosh (2\,\mu \,a\,r)-c_{0}},\hspace{0.5in%
}\mathrm{for}\hspace{0.5in}\lambda =\mu ,~c_{0}<-2.  \label{double1}
\end{equation}

\noindent and its warp factor is given by
\begin{equation}
e^{2\,A\left( r\right) }=N_{\alpha }\,\left[ \frac{2\,a}{\sqrt{c_{0}^{2}-4}%
\cosh (2\,\mu \,a\,r)-c_{0}}\right] ^{\left( \frac{4\,a^{2}}{9}\right) }\exp %
\left[ \frac{2\,a\left( c_{0}^{2}-c_{0}\,\sqrt{c_{0}^{2}-4}\cosh (2\,\mu
\,a\,r)-4\,a\right) }{9\left( \sqrt{c_{0}^{2}-4}\cosh (2\,\mu
\,a\,r)-c_{0}\right) ^{2}}\right] ,
\end{equation}

\noindent where, as we anticipated above, $N_{\alpha}$ will be chosen in
order to get $e^{2\,A\left( 0\right) }=1$, for plotting convenience.

An interesting aspect of these solutions is that, for some values of $%
c_{0}\,<-2$, $\tilde{\phi}_{DBW}^{(1)}(r)$ exhibits a double kink profile.
We can speak of a formation of a double wall structure, extended along the
space dimension. In the Fig.1 we compare the case studied in \cite{gomes} to
some typical profiles of the warp factors in the case where $\lambda =\mu $,
both when $c_{0}$ is close to its critical value ($c_{0}=-2$ in this case)
and far from it. One can observe the appearance of a more pronounced flat
region, where one could speak of a Minkowski-type metric. In fact this
\textquotedblleft Minkowski sector" becomes larger as $c_{0}$ approaches its
critical value.

Similar behavior is also noted in the classical solutions for $\lambda =4\mu
$ and $c_{0}<1/16$. In this case the field $\chi (r)$ has a lump-like
profile given by
\begin{equation}
\tilde{\chi}_{DBW}^{(2)}(r)=-\frac{2a}{\sqrt{\sqrt{1-16\, c_{0}}~\cosh
(4\,\mu \,a\,r)+1}},  \label{chi2atil2}
\end{equation}%
and the solution for the field $\phi (x)$ is
\begin{equation}
\tilde{\phi}_{DBW}^{(2)}(r)=\sqrt{1-16\,c_{0}}\,a\frac{\sinh (4\,\mu \,a\,r)%
}{\sqrt{1-16\,c_{0}}~\cosh (4\,\mu \,a\,r)+1},  \label{double2}
\end{equation}

\noindent with the corresponding warp factor being
\begin{eqnarray}
e^{2\,A\left( r\right) } &=&N_{\alpha }\,\left[ \frac{-\,2\,a}{\sqrt{\sqrt{%
1-16c_{0}}~\cosh (4\,\mu \,a\,r)+1}}\right] ^{\left( \frac{16\,\,a^{2}}{9}%
\right) }\times  \nonumber \\
&&\times \exp \left\{ -\frac{4\,a^{2}}{9}\left[ \frac{1+32\,a^{2}c_{0}+\sqrt{%
1-16c_{0}}~\cosh (4\,\mu \,a\,r)}{\left( \sqrt{1-16c_{0}}~\cosh (4\,\mu
\,a\,r)+1\right) ^{2}}\right] \right\} ,
\end{eqnarray}

\noindent and, once more we choose $N_{\alpha }$ in order to get $%
e^{2\,A\left( 0\right) }=1$.

In this last case $\tilde{\phi}_{DBW}^{(2)}(r)$ also presents a double kink
profile for some values of $c_{0}$. In Fig. 2 it can be seen the rising of
two peaks at the extremum of the flatten region and an increasing of the
distance between the two peaks in the warp factor as $c_{0}$ approaches its
critical value ($c_{0}$ = 1/16 in this case).

\subsection{Critical Bloch walls}

Finally, a very interesting class of analytical soliton solutions were shown
to exist when one takes $\lambda =\mu $ with the critical parameter $%
c_{0}=-2 $ and for $\lambda =4\mu $ with the critical parameter $c_{0}=1/16$%
, in the equation (\ref{eqdchi1}). The novelty in these cases is the fact
that both, the $\chi (r)$ and the $\phi (r)$ fields present a kink-like
profile and the warp factor presents a remarkable behavior, what can be
noted from Fig. 3, where it is evident the existence of two "Minkowski-type"
regions, separated by a transition one. This argument is reinforced by the
behavior of the energy densities of the soliton configurations, as can be
noted in the Fig. 4, as well as from the stability potential (see Fig. 5 and
6).

We call this set of solutions as $CBW$ domain walls. For $\lambda =\mu $ and
$c_{0}=-2$ the classical solution for the $\chi (r)$ field can be shown to
be
\begin{equation}
\chi _{CBW}^{\left( 1\right) }(r)=\frac{a}{2}\left[ 1\pm \tanh \left( \mu
\,a\,r\right) \right] ,  \label{chi2b}
\end{equation}%
and the solution for the $\phi (r)$ field is given by
\begin{equation}
\phi _{CBW}^{\left( 1\right) }(r)=\frac{a}{2}\left[ \tanh (\mu \;a\;r)\mp 1%
\right] .  \label{phi2b}
\end{equation}

\noindent The corresponding warp factor is
\begin{equation}
e^{2\,A\left( r\right) }=N_{\alpha }\,\left[ \frac{a}{2}\left[ 1\pm \tanh
\left( \mu \,a\,r\right) \right] \right] ^{\left( \frac{2\,\,a^{2}}{9}%
\right) }\exp \left\{ -\,\frac{a^{2}}{9}\left[ \left( 1\pm \tanh \left( \mu
\,a\,r\right) \right) ^{2}+\frac{c_{0}}{a}\left( 1\pm \tanh \left( \mu
\,a\,r\right) \right) \right] \right\} .
\end{equation}%
For $c_{0}=1/16$ and $\lambda =4\mu $, the following set of domain walls is
obtained
\begin{equation}
\tilde{\chi}_{CBW}^{\left( 2\right) }(x)=-\sqrt{2}a\frac{\cosh (\mu
\,a\,r)\pm \sinh (\mu \,a\,r)}{\sqrt{\cosh (2\,\mu \,a\,r)}},
\label{chi2btil}
\end{equation}%
and
\begin{equation}
\tilde{\phi}_{CBW}^{\left( 2\right) }(x)=\frac{a}{2}(1\mp \tanh (2\,\mu
\,a\,r)).  \label{phi2btil}
\end{equation}%
The warp factor, on its turn, is found to be given by
\begin{eqnarray}
e^{2\,A\left( r\right) } &=&N_{\alpha }\,\left[ \frac{-\,a\,e^{\pm \mu
\,a\,r}}{\sqrt{\cosh (2\,\mu \,a\,r)}}\right] ^{\left( \frac{16\,\,a^{2}}{9}%
\right) }\times  \nonumber \\
&& \\
&&\times \exp \left\{ -\,\frac{2\,a^{2}}{9}\tanh (2\,\mu \,a\,r)\left[
16\,a^{2}c_{0}\,\tanh \left( 2\,\mu \,a\,r\right) \mp (1+32\,a^{2}c_{0})%
\right] \right\} .  \nonumber
\end{eqnarray}

\section{Stability and zero modes}

In general it is quite hard to take into account a full set of fluctuations
of the metric around the background in a model where gravity is coupled to
scalars. This happens as a consequence of a very intricate system of coupled
non-linear second-order differential equations \cite%
{dewolfe,dewolfe1,gremm,gomes}. Fortunately however, there is a sector where
the metric fluctuations decouple from the scalars, and it comes to be the
one associated to the transverse and traceless part of the metric
fluctuation \cite{dewolfe,dewolfe1}. This can be shown, if one introduces a
metric perturbation like

\begin{equation}
ds^{2}=e^{2\,A\left( r\right) }\left( \eta _{\mu \nu }+\varepsilon \,h_{\mu
\nu }\right) dx^{\mu }dx^{\nu }-dr^{2},
\end{equation}

\noindent and perform small fluctuations on the scalar fields, $\phi
\rightarrow \phi \left( r\right) +\varepsilon ~\tilde{\phi}(r,x_{\mu })$ and
$\chi \rightarrow \chi \left( r\right) +\varepsilon ~\tilde{\chi}\left(
r,x_{\mu }\right) $, with $h_{\mu \nu }=h_{\mu \nu }\left( r,x_{\mu }\right)
$, and $\varepsilon $ is small perturbation parameter. Now, keeping the
terms in the action up to the second order in $\varepsilon $, as done
originally by DeWolfe and collaborators \cite{dewolfe}, as well as by Bazeia
and Gomes in the case of two scalar fields \cite{gomes}, one gets the
following set of coupled equations for the scalars fluctuations%
\begin{eqnarray}
e^{-2A}\square \tilde{\phi}-4\frac{dA}{dr}\frac{d\tilde{\phi}}{dr}-\frac{%
d^{2}\tilde{\phi}}{dr^{2}}+\frac{\partial ^{2}V}{\partial \phi ^{2}}~\tilde{%
\phi}+\frac{\partial ^{2}V}{\partial \phi \partial \chi }~\tilde{\chi} &=&%
\frac{1}{2}\frac{d\phi }{dr}\eta ^{\mu \nu }\frac{dh_{\mu \nu }}{dr},
\nonumber \\
&&  \label{flucteq} \\
e^{-2A}\square \tilde{\chi}-4\frac{dA}{dr}\frac{d\tilde{\chi}}{dr}-\frac{%
d^{2}\tilde{\chi}}{dr^{2}}+\frac{\partial ^{2}V}{\partial \chi ^{2}}~\tilde{%
\chi}+\frac{\partial ^{2}V}{\partial \phi \partial \chi }~\tilde{\phi} &=&%
\frac{1}{2}\frac{d\chi }{dr}\eta ^{\mu \nu }\frac{dh_{\mu \nu }}{dr},
\nonumber
\end{eqnarray}%
and, for the metric fluctuations, one obtains%
\[
-\frac{1}{2}\square h_{\mu \nu }+e^{2A}\left( \frac{1}{2}\frac{d}{dr}+2\frac{%
dA}{dr}\right) \frac{dh_{\mu \nu }}{dr}-\frac{1}{2}\eta ^{\alpha \beta
}\left( \partial _{\mu }\partial _{\nu }h_{\alpha \beta }-\partial _{\mu
}\partial _{\alpha }h_{\beta \nu }-\partial _{\nu }\partial _{\alpha
}h_{\beta \mu }\right) +
\]%
\begin{equation}
+\frac{1}{2}\eta _{\mu \nu }e^{2A}\frac{dA}{dr}\partial _{r}\left( \eta
^{\alpha \beta }h_{\alpha \beta }\right) +\frac{4}{3}e^{2A}\eta _{\mu \nu
}\left( \frac{\partial V}{\partial \phi }~\tilde{\phi}+\frac{\partial V}{%
\partial \chi }~\tilde{\chi}\right) =0.
\end{equation}%
One can simplify this last equation by choosing a transverse and traceless $%
h_{\mu \nu }$, which is done through the use of the projector%
\begin{equation}
P_{\mu \nu \alpha \beta }\equiv \frac{1}{2}\left( \pi _{\mu \alpha }~\pi
_{\nu \beta }+\pi _{\mu \beta }~\pi _{\nu \alpha }\right) -\frac{1}{3}\pi
_{\mu \nu }~\pi _{\alpha \beta },
\end{equation}%
where $\pi _{\mu \nu }\equiv \eta _{\mu \nu }-\frac{\partial _{\mu }\partial
_{\nu }}{\square }$. In other words, by using that $\bar{h}_{\mu \nu
}=P_{\mu \nu \alpha \beta }~h^{\alpha \beta }$, one arrives at
\begin{equation}
\frac{d^{2}\bar{h}_{\mu \nu }}{dr^{2}}+4\,\frac{dA}{dr}\,\frac{d\bar{h}_{\mu
\nu }}{dr}-e^{-2\,A}\partial _{\rho }\partial ^{\rho }\bar{h}_{\mu \nu }=0.
\label{eqST}
\end{equation}

\noindent Now, performing a sequence of function redefinition
\begin{equation}
\bar{h}_{\mu \nu }\equiv e^{i\,\vec{k}.\vec{x}}\,e^{-\frac{3}{2}\,A}\psi
_{\mu \nu },
\end{equation}

\noindent and variable transformation $z=\int e^{-A\left( r\right) }dr$ \cite%
{gomes}, one can recast the above equation into a kind of Schr\"{o}dinger
equation
\begin{equation}
-\frac{d^{2}\psi _{\mu \nu }}{dz^{2}}+U_{eff}\left( z\right) \psi _{\mu \nu
}=k^{2}\psi _{\mu \nu },
\end{equation}

\noindent where the effective potential is defined by%
\begin{equation}
U_{eff}\left( z\right) =\frac{9}{4}\left( \frac{dA}{dz}\right) ^{2}+\frac{3}{%
2}\frac{d^{2}A}{dz^{2}}.  \label{pz}
\end{equation}

\noindent The above differential equation can be factorized as
\begin{equation}
a^{+}\,a\,\psi _{\mu \nu }\left( z\right) \equiv \left( \frac{d}{dz}+\frac{3%
}{2}\frac{dA}{dz}\right) \left( -\frac{d}{dz}+\frac{3}{2}\frac{dA}{dz}%
\right) \psi _{\mu \nu }\left( z\right) =k^{2}\psi _{\mu \nu }.  \label{eqCT}
\end{equation}

\noindent It can be shown that $k^{2}$ is positive or zero since the
resulting Hamiltonian can be factorized as the product of two operators wich
are adjoints of each other \cite{bazeiaPLB06}. So the system is stable
against linear classical metric fluctuations.

Regarding the stability of the system against classical linear fluctuations
of the scalar fields we note that the equations (\ref{flucteq}) are quite
hard to be analyzed due to the coupling of the fields among themselves and
with the metric fluctuations (The term in the right hand side of the
equations (\ref{flucteq})). Thus, one can try to simplify the problem by
considering only the fluctuations of one of the scalar fields. If we
consider, for instance, only the fluctuation of $\ \chi (r)$, we are left
just with the second of the equations (\ref{flucteq}). Performing the field
transformation $\tilde{\chi}(r)=e^{i\,\vec{p}.\vec{x}}\,e^{-\frac{3}{2}%
\,A}\zeta (z)$ we obtain
\begin{equation}
-\frac{d^{2}\zeta }{dz^{2}}+\left( U_{eff}\left( z\right) +e^{2A(z)}\frac{%
\partial ^{2}V}{\partial \chi ^{2}}\right) \zeta =p^{2}\zeta ,
\label{eqlump}
\end{equation}%
where $U_{eff}\left( z\right) $ is given in the equation (\ref{pz}) and we
have taken into account only the zero-mode of the metric fluctuation. This
can be done as far as the potential $U_{eff}\left( z\right) \,$, the one
responsible to localize the gravity, supports only the zero mode as a
localized state. Unfortunately, we have not been able to factorize the
equation (\ref{eqlump}) as a product of two operators wich are adjoint to
each other, as in equation (\ref{eqCT}). If that is possible the system is
also stable against the fluctuations of at least one of the scalar fields.
Furthermore, in general the factor $e^{2A(z)}(\partial ^{2}V/\partial \chi
^{2})$ is not positive for all values of the variable $z$, thus we can not
guarantee that the spectrum is positive semi-definite. As far as we know the
question regarding the stability of the scalar fields is an open problem. In
the reference (\ref{dewolfe1}) this question was thoroughly examined for the
situation where only one active scalar field is present and all the metric
fluctuations modes were fully taken into account. The authors show that in
some cases the stability of domain walls can be proven, that is $p^{2}\geq 0$%
, although the effective potential can not be factorized. Unfortunately, the
case we are studying here does not belongs to any of those cases.

Returning to the analysis of the metric fluctuations equation, we remark
that the zero mode coming from the equation (44) grants the existence of
massless four-dimensional gravitons \cite{randall, gremm, furtado,gomes}. In
general the shape of the zero mode is quite similar to the warp factor so
that one can think that there is some relation between them. In the next we
will show that these two quantities really present the same generic shape.
With this in mind, we start from equation (\ref{eqST}), redefine the
function $\bar{h}_{\mu \nu }$ as $\bar{h}_{\mu \nu }=e^{-2\,A\left( r\right)
}\xi _{\mu \nu }\left( r\right) $, and obtain the following equation
\begin{equation}
-\frac{d^{2}\xi _{\mu \nu }}{dr^{2}}+2\left( \frac{d^{2}A}{dr^{2}}+2\left(
\frac{dA}{dr}\right) ^{2}\right) \xi _{\mu \nu }-k^{2}\,e^{-2\,A}\xi _{\mu
\nu }=0,
\end{equation}

\noindent which, for the case of the zero mode ($k^{2}=0$) can be rewritten
as
\begin{equation}
\left( \frac{d}{dr}+2\,\frac{dA}{dr}\right) \left( -\frac{d}{dr}+2\,\,\frac{%
dA}{dr}\right) \xi _{\mu \nu }=0,
\end{equation}

\noindent and finally one can see that the zero mode solution, apart from a
normalization factor, is precisely the warp factor
\begin{equation}
\xi _{\mu \nu }^{\left( 0\right) }=N_{0}\,e^{2\,A\left( r\right) }\eta _{\mu
\nu }.
\end{equation}

In terms of the coordinate $r$ the effective potential which localize the
gravitation in the brane is written as
\begin{equation}
U_{eff}\left( r\right) =\frac{3}{4}e^{2\,A}\,\,\left( 2\,\frac{d^{2}A}{dr^{2}%
}+5\,\left( \frac{dA}{dr}\right) ^{2}\right) .
\end{equation}

\noindent Obviously the above potential is equal to the one in the $z$
variable, it will be a kind of re-scaled one. However the general shape and
characteristics in both variables are the same. The stability potential is
represented in figures 5-7. In Fig. 5 we compare the behavior of one of our
degenerate cases with those of Bazeia and Gomes \cite{gomes}. Fig. 6 shows
clearly that the structure of the potential in a situation where two
interactive regions are separated by an approximately zero force one.
Finally, in Fig. 7, we see that those separated potentials recombine into a
single one.

The essential idea in our work is to show that the situation is much richer
than that analyzed in \cite{gomes}, and that from a complete set of
solutions as the one we present here, important consequences for the warp
factor structure and, consequently, for the brane world scenario show up.
One can cite for instance the fact that one can control, by means of a
parameter which is not present in the potential $V\left( \phi ,\chi \right) $%
, the region where the metric is approximately flat. Furthermore, in a given
critical case, there are two of these regions, separated by a transition one
(see Fig. 3).

\bigskip

\section{Final remarks}

In this work we analyze the impact of a general set of solitonic solutions
over the characteristics of some models presenting interaction between two
scalar fields coupled to gravity in (4,1) dimensions in warped space-time
with one extra dimension. Essentially, we explore a larger class of
solutions of a model recently studied \cite{gomes}. In doing so, we have
discovered a number of interesting features as, for instance, a kind of
type-I extreme domain wall as classified in \cite{cvetic}, when we have
dealt with what we have called critical domain walls (see Fig. 3).

One very important consequence of our study is that the thickness of the
domain walls can be controlled by means of an external parameter (regarding
the scalar fields potential), and this can be done without changing the
potential parameters, in contrast with is done in other models \cite%
{gomes,furtado}.

Furthermore, one can observe the appearance of a controllable flat region in
the warp factor, where one could speak of a Minkowski-type metric region
(see Fig. 2 and 4). In fact, in Fig. 4, where the energy density is plotted,
we see clearly in the case with $c_{0}=-2.00001$, that the region where
negative energy densities show up is outside of the ''Minkowskian'' one.
Thus one could speculate about a possible confining mechanism for the bulk
particles in that internal region.

\textbf{Acknowledgements: }The authors ASD and MH thanks to CNPq and ACAF to
CAPES for the partial financial support. We also thanks to Professor D.
Bazeia for introducing us to the matter of solitons and BPS solutions. This
work has been finished during a visit (ASD) within the Associate Scheme of
the Abdus Salam ICTP.

\newpage

\begin{figure}[tbp]
\begin{center}
\begin{minipage}{20\linewidth}
\epsfig{file=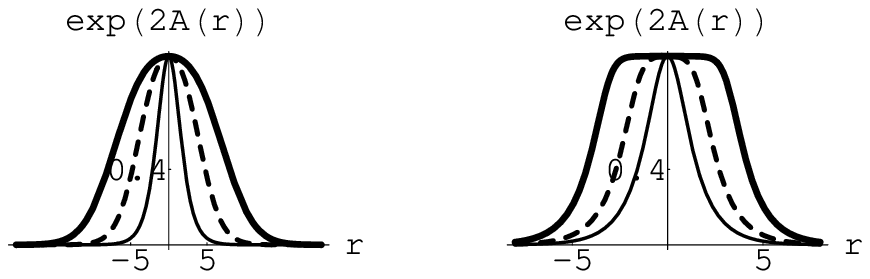}
\end{minipage}
\end{center}
\caption{Warp factor appearing in the reference \protect\cite{gomes}, with
the parameters used there: $r=0.05$ (thick solid line), $0.1$ (dashed line)
and $0.3$ (thin solid line) (left). Warp factor for the case where: $a=1$ $%
\protect\lambda =\protect\mu$, $c_0\neq -2$ and $c_0=-2.00001$ (thick solid
line); $-2.005$ (dashed line); $-3.0$ (thin solid line) (right).}
\label{fig:fig1}
\end{figure}

\begin{figure}[tbp]
\begin{center}
\begin{minipage}{20\linewidth}
\epsfig{file=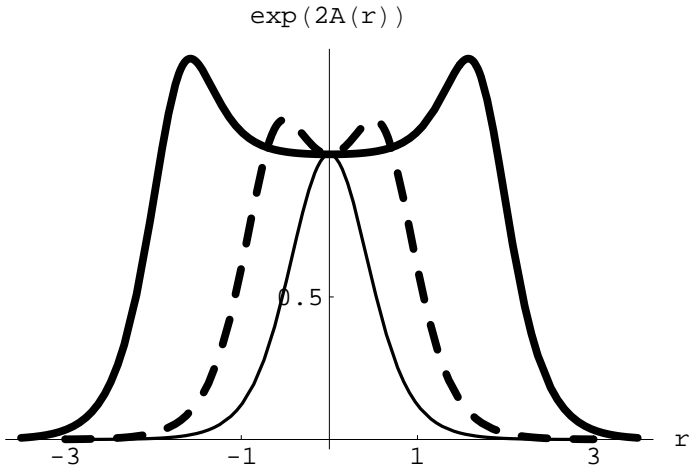}
\end{minipage}
\end{center}
\caption{Warp factor for the case where: $a=1$ $\protect\lambda =4 \protect%
\mu$, $c_0\neq 1/16$ and $c_0=1/16.0001$ (thick solid line); $1/17$ (dashed
line); $1/200$ (thin solid line). Note the appearance of the two peaks,
signalizing a richer structure for the zero mode.}
\label{fig:fig2}
\end{figure}

\begin{figure}[tbp]
\begin{center}
\begin{minipage}{20\linewidth}
\epsfig{file=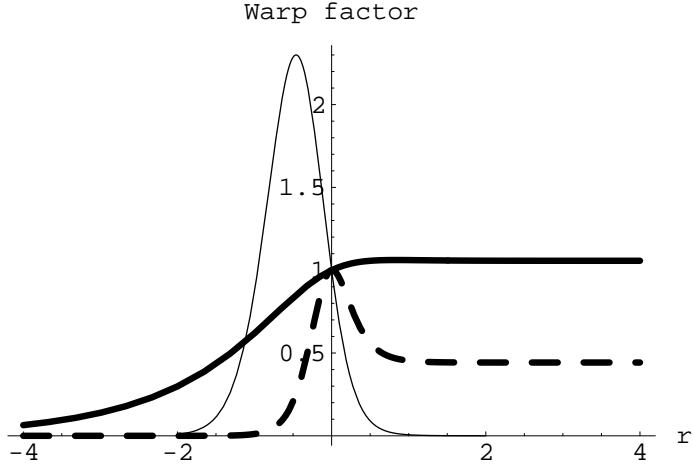}
\end{minipage}
\end{center}
\caption{Warp factor for the case where: $\protect\lambda =4 \protect\mu$, $%
c_0= 1/16$, $\protect\mu =1$, $a=0.6$ (thick solid line), $a=1.2$ (dashed
line) and the case where $a=2$ and $\protect\mu=0.2$.}
\label{fig:fig3}
\end{figure}

\begin{figure}[tbp]
\begin{center}
\begin{minipage}{20\linewidth}
\epsfig{file=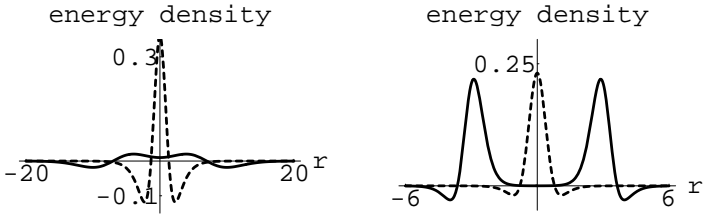}
\end{minipage}
\end{center}
\caption{Energy density of reference \protect\cite{gomes} $r=0.05$ and $0.30$
(left). Energy density for the case where: $a=1$ $\protect\lambda =\protect%
\mu=1$, $c_0\neq -2$ and $c_0=-2.00001$ (solid line); $-4.0$ (dashed line)
(right).}
\label{fig:fig4}
\end{figure}

\begin{figure}[tbp]
\begin{center}
\begin{minipage}{20\linewidth}
\epsfig{file=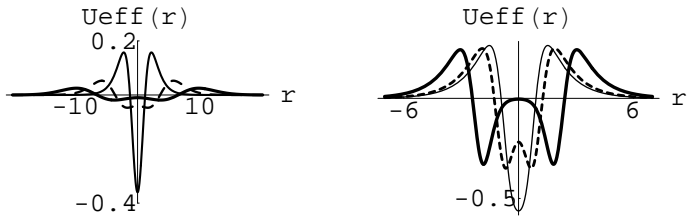}
\end{minipage}
\end{center}
\caption{Comparison of the stability potential of reference \protect\cite%
{gomes} (left) with $r=0.05; 0.1; 0.3$ and ours with $\protect\lambda =%
\protect\mu$, $a=1$ and $c_0=-2.001$ (thick solid line); $-2.1$ (dashed
line); $-2.5$ (thin solid line).}
\label{fig:fig5}
\end{figure}

\begin{figure}[tbp]
\begin{center}
\begin{minipage}{20\linewidth}
\epsfig{file=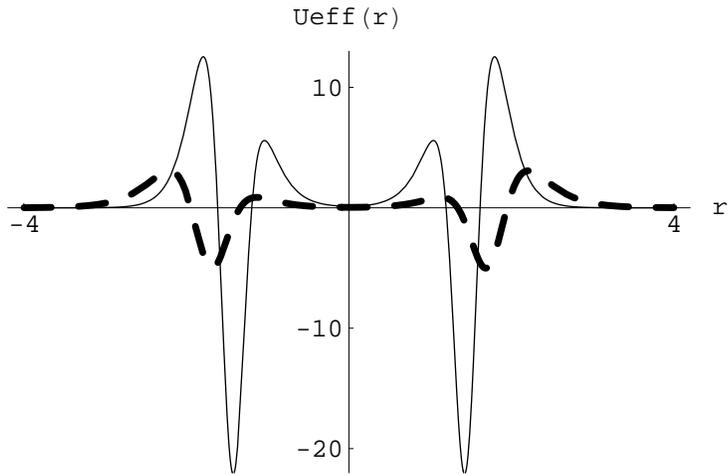}
\end{minipage}
\end{center}
\caption{Stability potential for the case where $c_0=1/16.0001$ with $a=1$
(thin dashed line), and $a=1.2$ (thick solid line)}
\label{fig:fig6}
\end{figure}

\begin{figure}[tbp]
\begin{center}
\begin{minipage}{20\linewidth}
\epsfig{file=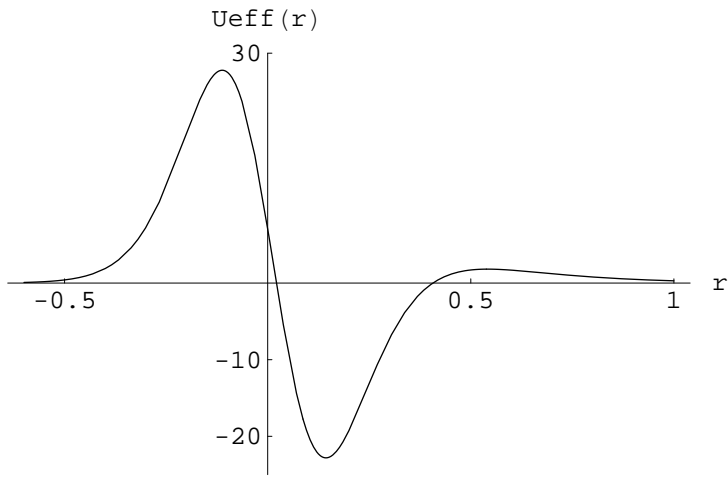}
\end{minipage}
\end{center}
\caption{A typical stability potential for the critical case, both when $%
\protect\lambda =\protect\mu$ as $\protect\lambda =4\,\protect\mu$.}
\label{fig:fig7}
\end{figure}

\end{document}